\documentclass[twocolumn,prb,showpacs]{revtex4}

\usepackage{graphicx}
\usepackage{dcolumn}
\usepackage{bm}

\begin{document}

\title{Critical fields and spontaneous vortex state in the weak-ferromagnetic superconductor RuSr$_2$GdCu$_{2}$O$_{8}$}

\author{C. Y. Yang}
\author{B. C. Chang}
\author{H. C. Ku}%
 \email{hcku@phys.nthu.edu.tw}
 \affiliation{Department of Physics, National Tsing Hua University, Hsinchu, Taiwan 300, R.O.C.}
\author{Y. Y. Hsu}
\affiliation{Institute of Physics, Academia Sinica, Taipei, Taiwan 115, R.O.C.}

\date{\today}
                                                                
                                             
\begin{abstract}
A spontaneous vortex state (SVS) between 30 K and 56 K was observed for the weak-ferromagnetic 
 superconductor RuSr$_{2}$GdCu$_{2}$O$_{8}$ with ferromagnetic Curie temperature T$_{C}$ = 131 K and superconducting transition temperature T$_{c}$ = 56 K. The low field ($\pm$20 G) superconducting hysteresis loop indicates a narrow Meissner state region within average lower critical field B$_{c1}$(T)= B$_{c1}$(0)[1 - (T/T$_{0}$)$^{2}$], with average B$_{c1}^{ave}$(0) = 12 G and T$_{0}$ = 30 K. Full Meissner shielding signal in very low applied field indicates an ab-plane B$_{c1}^{ab}$(0) $\sim$ 4 G with an estimated anisotropic parameter $\gamma \sim$ 7 for this layered system. The existence of a spontaneous vortex state between 30 K and 56 K is the result of weak-ferromagnetic order with a net spontaneous magnetic moment of $\sim$ 0.1 $\mu_{B}$/Ru, which generates a weak magnetic dipole field around 10 G in the CuO$_{2}$ bi-layers. The upper critical field B$_{c2}$ varies linearly as (1 - T/T$_{c}$) up to 7-T field. The vortex melting line B$_{m}$ varies as (1 - T/T$_{m}$)$^{3.5}$ with melting transition temperature T$_{m}$ = 39 K and a very broad vortex liquid region due to the coexistence and the interplay between superconductivity and weak-ferromagnetic order. 
\end{abstract}

\maketitle

\section{Introduction}

Recently, high-T$_{c}$ superconductivity with anomalous magnetic properties was reported in the weak-ferromagnetic Ru-1212 system 
RuSr$_{2}$RCu$_{2}$O$_{8}$ (R = Sm, Eu, Gd, Y) with the tetragonal TlBa$_{2}$CaCu$_{2}$O$_{7}$-type structure.\cite{p1,p2,p3,p4,p5,p6,p7,p8,p9,p10,p11,p12,p13,p14,p15,p16,p17,p18,p19,p20,p21,p22,p23,p24,p25,p26,p27,p28,p29,p30,p31,p32,p33,p34,p35,p36,p37,p38,p39,p40,p41,p42,p43} For the Ca-substituted system, a possible superconductivity was also reported in the weak-ferromagnetic compounds RuCa$_{2}$RCu$_{2}$O$_{8}$ (R = Pr-Gd).\cite{p44,p45,p46} 
The metallic weak-ferromagnetic (WFM) order is originated from the long range order of Ru 
moments in the RuO$_{6}$ octahedra due to strong Ru-4d$_{xy,yz,zx}$-O-2p$_{x,y,z}$ hybridization in this strongly-correlated electron system. The Curie temperature T$_{C} \sim$ 130 K observed from magnetization measurement in the prototype compound RuSr$_{2}$GdCu$_{2}$O$_{8}$ is probably a canted G-type antiferromagnetic order with Ru$^{5+}$ moment $\mu$ canted along the tetragonal basal plane resulting a small net spontaneous magnetic moment $\mu_{s}$ $\ll \mu$(Ru$^{5+}$) too small to be detected in neutron diffraction.\cite{p4,p5,p9,p10,p21} The occurrence of high-T$_{c}$ superconductivity with maximum resistivity onset T$_{c}$(onset) $\sim$ 60 K in RuSr$_{2}$GdCu$_{2}$O$_{8}$ is related with the quasi-two-dimensional CuO$_{2}$ 
bi-layers separated by a rare earth layer in the Ru-1212 structure.\cite{p1,p2,p4,p5,p29} Broad resistivity transition width $\Delta$T$_{c}$ = T$_{c}$(onset) - T$_{c}$(zero) $\sim$ 15-20 K observed is most likely originated from the coexistence and the interplay between superconductivity and weak-ferromagnetic order.\cite{p1,p2,p3,p4,p5,p6,p7,p8,p9,p10,p11,p12,p13,p14,p15,p16,p17,p18,p19,p20,p21,p22,p23,p24,p25,p26,p27,p28,p29,p30,p31,p32,p33,p34,p35,p36,p37,p38,p39,p40,p41,p42,p43} The diamagnetic T$_{c}$ is observed anomalously at lower temperature near T$_{c}$(zero) instead of at T$_{c}$(onset), and a reasonable large Meissner signal was reported using stationary sample magnetometer with diamagnetic T$_{c} \sim$ 30 K in $\le$ 1 G applied field at zero-field-cooled (ZFC) mode.\cite{p38} Lower T$_{c}$(onset) $\sim$ 40 K and 12 K were observed for RuSr$_{2}$EuCu$_{2}$O$_{8}$ and RuSr$_{2}$SmCu$_{2}$O$_{8}$, respectively.\cite{p12,p17} No superconductivity can be detected in RuSr$_{2}$RCu$_{2}$O$_{8}$ (R = Pr, Nd).\cite{p3,p16} Superconducting RuSr$_{2}$YCu$_{2}$O$_{8}$ phase is stable only under the high pressure.\cite{p20,p25}

The physics is still unclear in this system, and it will be interesting to investigate the effect of the weak-ferromagnetic order on superconducting critical fields B$_{c2}$ and B$_{c1}$, as well as on the possible existence of a spontaneous vortex state (SVS) at higher temperature above the Meissner state.
 
\section{experimental}
The stoichiometric RuSr$_{2}$GdCu$_{2}$O$_{8}$ samples were 
synthesized by the standard solid-state reaction method. High-purity RuO$_{2}$ (99.99 $\%$), 
SrCO$_{3}$ (99.99 $\%$), Gd$_{2}$O$_{3}$ (99.99 $\%$), and CuO (99.99 $\%$) preheated powders with
the nominal composition ratio of Ru:Sr:Gd:Cu = 1: 2: 1: 2 were well mixed and calcined at 960$^{\circ}$C in air for 16 hours. The calcined powders were then pressed into pellets and sintered 
in flowing N$_{2}$ gas at 1015$^{\circ}$C for 10 hours to form RuSr$_{2}$GdO$_{6}$ and Cu$_{2}$O 
precursors. This step is crucial in order to avoid the formation of unwanted impurity phases. The N$_{2}$-sintered pellets were heated at 1060$^{\circ}$C in flowing 
O$_{2}$ gas for 10 hours to form the Ru-1212 phase. The pellets were oxygen-annealed at slightly higher 
1065$^{\circ}$C for 5 days and slowly furnace-cooled to room temperature with a rate of 
15$^{\circ}$C per hour.\cite{p15}

The powder x-ray diffraction data were collected with a Rigaku Rotaflex 
18-kW rotating anode diffractometer using graphite monochromatized Cu-K$_{\alpha}$ radiation 
with a scanning step of 0.02$^{\circ}$ (10 second counting time per step) in the 2 $\theta$ 
ranges of 5-100$^{\circ}$. The electrical resistivity and magneto-resistivity measurements were performed using the standard 
four-probe method with a Linear Research LR-700 ac (16Hz) resistance bridge from 2 K to 300 K with 
applied magnetic field up to 7 T. The magnetization, magnetic susceptibility and magnetic hysteresis measurements from 2 K to 
300 K with applied fields from 1 G to 7 T were carried out with a Quantum Design 1-T 
$\mu$-metal shielded MPMS2 or a 7-T MPMS superconducting quantum interference device (SQUID) 
magnetometer.

\begin{figure}
\includegraphics{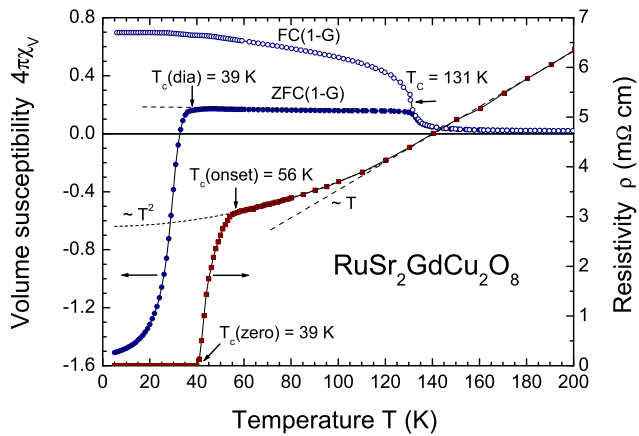}
\caption{\label{fig:epsart} Electrical resistivity $\rho$(T) and volume magnetic susceptibility 
$\chi_{V}$(T) at 1-G field-cooled (FC) and zero-field-cooled (ZFC) modes for oxygen-annealed 
RuSr$_{2}$GdCu$_{2}$O$_{8}$}
\end{figure}

\section{results and discussion}
The powder x-ray diffraction pattern for the oxygen-annealed RuSr$_{2}$GdCu$_{2}$O$_{8}$ polycrystalline sample
indicates close to single phase with the tetragonal lattice parameters of $\it{a}$ = 0.5428(5) nm 
and $\it{c}$ = 1.1589(9) nm. The space group P4/mbm is used for Rietveld refinement analysis, where neutron diffraction data indicate that a RuO$_{6}$ octahedra 14$^{\circ}$ rotation around the c-axis is needed to accommodate physically reasonable Ru-O bond lengths.\cite{p10} The refinement gives a good residual error 
R of 3.64 $\%$, weighted pattern error R$_{wp}$ = 6.07 $\%$, and Bragg error R$_{B}$ = 5.05 $\%$.

The temperature dependence of the electrical resistivity $\rho$(T) and the volume 
magnetic susceptibility $\chi_{V}$(T) at 1-G field-cooled (FC) and zero-field-cooled (ZFC) 
modes for RuSr$_{2}$GdCu$_{2}$O$_{8}$ are shown collectively in Fig. 1. The high temperature resistivity 
decreases monotonically from room temperature value of 9.2 m$\Omega$.cm (not shown) to 6.4 m$\Omega$.cm 
at 200 K, and extrapolated to 2.8 m$\Omega$.cm at 0 K with a good resistivity ratio $\rho$(300 K)/$\rho$(0 K) 
of 3.3 for the polycrystalline sample. High temperature resistivity shows a non-Fermi-liquid-like linear T-dependence 
down to Curie temperature T$_{C}$ of 131 K, then changes to a T$^{2}$ behavior below T$_{C}$ due to magnetic order.

The superconducting onset temperature of 56 K is determined from 
the deviation from T$^{2}$ behavior, with a zero resistivity T$_{c}$(zero) at 39 K. The broad transition width $\Delta$T$_{c}$ = 17 K observed is the common feature for all reported Ru-1212 resistivity data, which indicates that the superconducting Josephson coupling along the tetragonal c-axis 
between Cu-O bi-layers may be partially blocked by the dipole field B$_{dipole}$ of ordered Ru moments in the Ru-O 
layer.\cite{p1,p2,p4,p5,p29,p40} The diamagnetic T$_{c}$ at 39 K was observed in the 1-G ZFC susceptibility 
measurement. The full Meissner shielding signal 4$\pi\chi_{V}$ = 4$\pi$M/B$_{a} \sim$ -1.5 (Gaussian units) was recorded at 5 K. This value is identical 
to the Meissner shielding signal expected for a superconducting sphere with a demagnetization factor N of -4$\pi$/3 and in an applied field B$_{a}$ well below lower critical field B$_{c1}$. The large diamagnetic signal 
in 1-G ZFC mode is the best data observed so far from various reported susceptibility measurement techniques.\cite{p4,p5,p28,p29,p38} Since our measurements were performed with the standard moving-sample SQUID magnetometer, it is clear that sample quality is more crucial than measuring techniques. Both ZFC and FC data reveal a Curie temperature T$_{C}$ of 131 K. However, in 1-G FC mode, no diamagnetic field-expulsion signal can be detected below 39 K due to strong flux pinning where superconductivity coexists with weak-ferromagnetic order.

\begin{figure}
\includegraphics{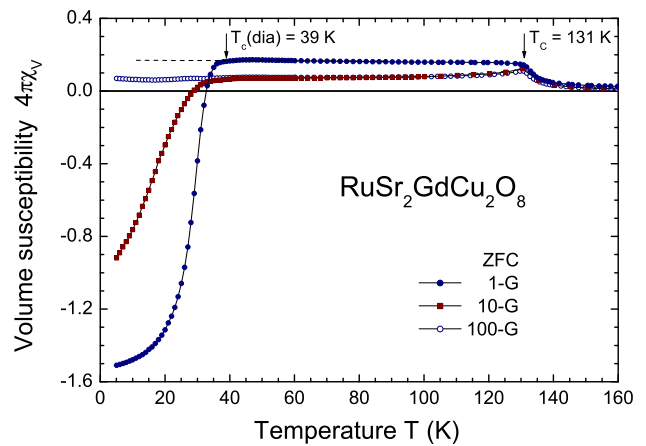}
\caption{\label{fig2} ZFC volume susceptibility $\chi_{V}$(T) for RuSr$_{2}$GdCu$_{2}$
O$_{8}$ at 1 G, 10 G, and 100 G. Note that the full Meissner shielding signal was observed only at low 
applied field and low temperature.}
\end{figure}

The zero-field-cooled (ZFC) volume susceptibility $\chi_{V}$(T) at 1 G, 10 G, and 100 G applied 
fields are shown collectively in Fig. 2. All data show the same magnetic order T$_{C}$(Ru) of 131 K. Although the diamagnetic T$_{c}$ of 39 K was still 
observed at 10-G ZFC measurement, the diamagnetic signal at 5 K is reduced to 60$\%$ of the full 
Meissner signal. Consider the polycrystalline nature of sample with varying microcrystallite size and orientation, the average superconducting lower critical field B$_{c1}$ at 5 K is estimated to be close to 10 G. No net diamagnetic signal can be detected at 100-G ZFC mode where the sample is already in the vortex glass or lattice state and the small diamagnetic signal is overshadowed by a large weak-ferromagnetic background.\cite{p38}

\begin{figure}
\includegraphics{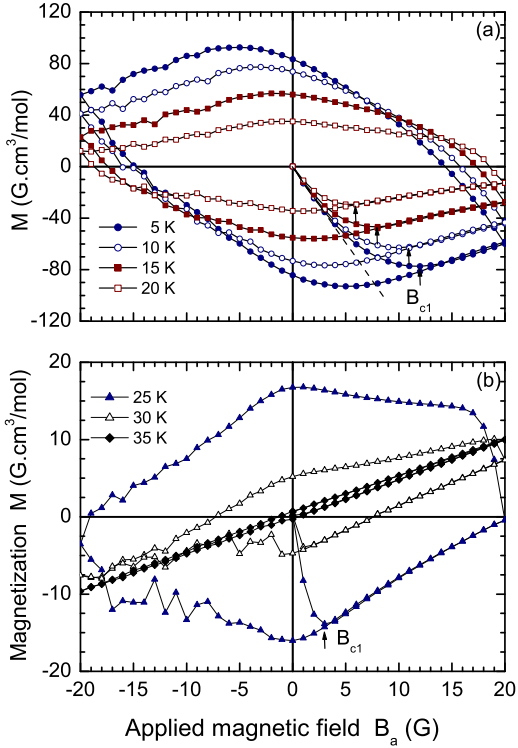}
\caption{\label{fig3}  Low field superconducting hysteresis loops M-B$_{a}$ for 
RuSr$_{2}$GdCu$_{2}$O$_{8}$: (a) at 5 K, 10 K, 15 K, 20 K; (b) at 25 K, 30 K, and 35 K.}
\end{figure}

Based on this information, the low-field ($\pm$20 G) isothermal superconducting hysteresis loops
M-B$_{a}$ are measured and collectively shown in Fig. 3(a) (5 K, 10 K, 15 K, and 20 K) and 3(b) (25 K, 30 K, and 35 K). The initial magnetization curve deviates from straight line in 4 G at 5 K, 3.5 G at 10 K, 3 G at 15 K, 2 G at 20 K, and 1 G at 25 K. This is the narrow region that full Meissner signals are detected and is roughly corresponding to the anisotropic lower critical field in the ab-plane B$_{c1}^{ab}$(T) with B$_{c1}^{ab}$(0) $\sim$ 4 G. The average lower critical field B$_{c1}^{ave}$ for polycrystalline sample is determined from the peaks of initial diamagnetic magnetization curves. B$_{c1}$ decreases steadily from 12 G at 5 K, 11 G at 10 K, 9 G at 15 K, 6 G at 20 K, 3 G at 25 K and below 1 G at 30 K. A simple empirical parabolic fitting gives B$_{c1}$(T) = B$_{c1}$(0)[1 - (T/T$_{0}$)$^{2}$], with average B$_{c1}^{ave}$(0) = 12 G and 
T$_{0}$ = 30 K (see Fig. 4). Using the anisotropic Ginzburg-Landau formula B$_{c1}^{ave}$ = [2B$_{c1}^{ab}$ + B$_{c1}^{c}$]/3, c-axis B$_{c1}^{c} \sim$ 28 G and anisotropy parameter $\gamma \sim$ 7 is estimated. This value is close to reported anisotropic $\gamma$-value for YBa$_{2}$Cu$_{3}$O$_{7}$ where the 123-type structure can be written as Cu-1212 CuBa$_{2}$YCu$_{2}$O$_{7}$. An average penetration depth $\lambda_{ave}$(0) = [$\Phi_{0}$/2$\pi$B$_{c1}^{ave}$(0)]$^{1/2}$ of 520 nm was derived with estimated $\lambda_{ab}$(0) = 340 nm and $\lambda_{c}$(0) = 2400 nm from B$_{c1}^{c}$ = $\Phi_{0}$/2$\pi\lambda_{ab}^{2}$ and B$_{c1}^{ab}$ = $\Phi_{0}$/2$\pi\lambda_{ab}\lambda_{c}$, where $\Phi_{0}$ is flux quantum.

\begin{figure}
\includegraphics{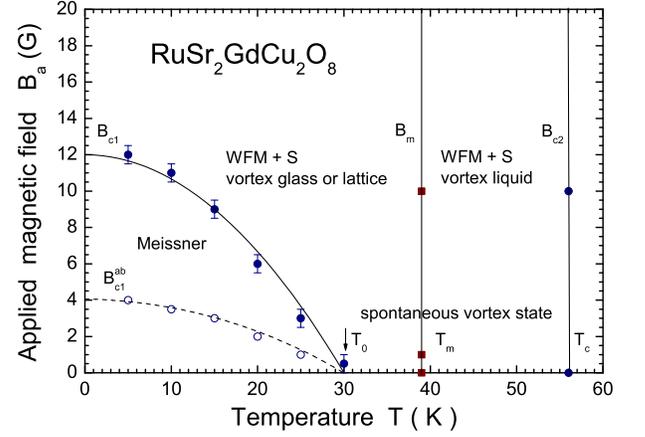}
\caption{\label{fig4}  The lower field, low temperature superconducting phase diagram B$_{a}$(T) of RuSr$_{2}$GdCu$_{2}$O$_{8}$.}
\end{figure}

Since T$_{0}$ = 30 K is well below T$_{c}$(onset) = 56 K and T$_{c}$(zero) = 39 K in zero applied field, a spontaneous vortex state (SVS) indeed exists between 30 K and 56 K. The low field phase diagram B$_{a}$(T) for polycrystalline sample is shown in Fig. 4, with the average B$_{c1}$(T) separates the Meissner state from the vortex state and a smaller B$_{c1}^{ab}$(T) inside the Meissner region for reference. T$_{c}$(zero) = 39 K in the broad resistive transition is the onset of vortex depinning by driving current. This temperature is very close to the melting transition temperature T$_{m}$ from the spontaneous vortex glass or lattice state to the spontaneous liquid state due to nonzero dipole field B$_{dipole}$ of weak-ferromagnetic order. The upper critical field B$_{c2}$ defined from T$_{c}$(onset) and the vortex melting field B$_{m}$(T) defined from T$_{c}$(zero) are temperature independent for small applied fields below 20 G. The internal dipole field generated by a weak-ferromagnetic order can be estimated using a simple extrapolation [B$_{c1}$(0)+ B$_{dip}$]/B$_{c1}$(0) = T$_{c}$/T$_{0}$ = 56 K/30 K, which results with a dipole field B$_{dipole}$ $\sim$ 10.4 G on the CuO$_{2}$ bi-layers. A small net spontaneous magnetic moment $\mu_{s}$ of $\sim$ 0.11 $\mu_{B}$ per Ru is estimated using B$_{dipole} \sim$ 2$\mu_{s}$/d$^{3}$ with d = $\it{c}$/2 = 0.58 nm is the distance between midpoint of CuO$_{2}$ bi-layers and two nearest-neighbor Ru moments. If the weak-ferromagnetic structure is a canted G-type antiferromagnetic order with Ru moments $\mu$ (= 1.5 $\mu_{B}$ for Ru$^{5+}$ in t$_{2g}$ states) canted along the tetragonal basal plane, the small net spontaneous magnetic moment gives a canting angle of 4$^{\circ}$ from the tetragonal c-axis and is difficult to be detected in neutron diffraction with a resolution around 0.1 $\mu_{B}$.\cite{p9,p10,p21}

At 5 K, the shape of superconducting hysteresis loop with a large remanent molar magnetization M$_{r}$ of 83 G.cm$^{3}$/mol indicates strong pinning as well as a good indication of bulk nature of superconductivity for the oxygen-annealed sample. The remanent M$_{r}$ decreases to 4 G.cm$^{3}$/mol at 30 K and 1 G.cm$^{3}$/mol at 35 K, where a weak-ferromagnetic background can be clearly seen. Fluctuation in the hysteresis loop is probably also related to the weak-ferromagnetic order.

\begin{figure}
\includegraphics{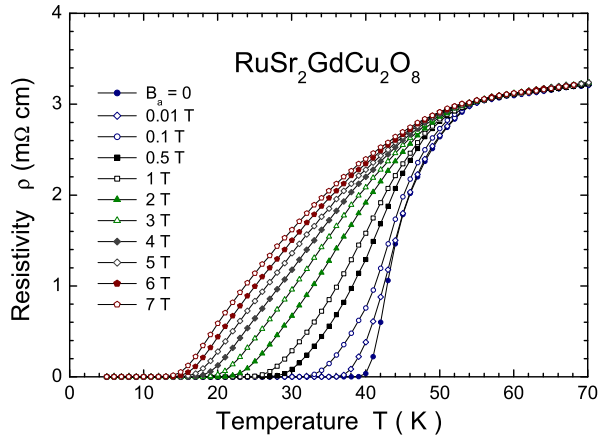}
\caption{\label{fig5}  Temperature dependence of magnetoresistivity $\rho$(T, B$_{a}$) for 
RuSr$_{2}$GdCu$_{2}$O$_{8}$ in applied field up to 7 T.}
\end{figure}

To study the high-field effect on superconductivity, the magnetoresistivity $\rho$(T,B$_{a}$) for RuSr$_{2}$GdCu$_{2}$O$_{8}$ up to 7 T are collectively 
shown in Fig. 5. The broadening of resistive transition in magnetic fields is the common features for all high-T$_{c}$ cuprate superconductors.\cite{p47} The normal state resistivity is field independent and follows a T$^{2}$-dependence below T$_{C}$, with superconducting T$_{c}$(onset) of 56 K in zero field decreases slightly to 53 K in 7-T field. The temperature dependence of upper critical field B$_{c2}$(T) can be fitted with a linear function B$_{c2}$(0)[1 - T/T$_{c}$] with average B$_{c2}$(0) = 133 T.\cite{p47} An average coherence length $\xi_{0}^{ave}$ = [$\Phi_{0}$/2$\pi$B$_{c2}^{ave}$(0)]$^{1/2}$ of 0.5 nm with the Ginzburg-Landau parameter $\kappa$ of 1040 and the thermodynamic critical field B$_{c}$(0) = (B$_{c1}$.B$_{c2}$)$^{1/2}$ = 0.32 T. No anisotropic $\xi_{ab}$ and $\xi_{c}$ values can be estimated from present data. The T$_{c}$(zero) decreases from 39 K in zero applied field to 32 K in 1-kG, 28 K in 5-kG, 25 K in 1-T, 22 K in 2-T, 19 K in 3-T, 17 K in 4-T, 16 K in 5-T, 15 K in 6-T, and 14 K in 7-T field. If the zero resistivity is taken as the lower bound of the vortex melting temperature T$_{m}$, then the temperature dependence of the vortex melting transition line B$_{m}$(T) can be fitted roughly by the formula B$_{m}$(T) = B$_{m}$(0)[1 - T/T$_{m}$]$^{3.5}$ with B$_{m}$(0) = 35 T and large exponent 3.5. In the lower field region, B$_{m}$(T) rises as [1 - T/T$_{m}$]$^{2}$ as predicted by the mean-field approximation for temperature near T$_{m}$ = 39 K.\cite{p47} The full phase diagram B$_{a}$(T) of RuSr$_{2}$GdCu$_{2}$O$_{8}$ is shown in Fig. 6 to exhibit both high field and low field features. The very broad vortex liquid region with $\Delta$T = 17 K in zero field and $\Delta$T = 42 K in 7-T field is extraordinary and is most likely originated from the coexistence and the interplay between superconductivity and weak-ferromagnetic order. This magnetic order is so weak that superconductivity can coexist with the magnetic order, but the effect of a weak spontaneous magnetic moment $\mu_{s} \sim$ 0.1 $\mu_{B}$ is detected through the appearance of a spontaneous vortex state above 30 K with a broad spontaneous vortex liquid region above T$_{m}$ of 39 K.

To study the broad vortex liquid region, the isothermal field-dependent magnetoresistivity $\rho$(B$_{a}$) for T $<$ T$_{c}$ are shown in Fig. 7, where zero resistivity gives a lower bound of vortex melting field B$_{m}$. In the resistive vortex liquid region, there is no unique power-law fit, but the data can be roughly fitted with various powers between 1/6 and 1, depending on temperature range. At T = 40 K $>$ T$_{m}$ = 39 K, resistivity is $\sim$ B$_{a}^{1/6}$ for B$_{a} >$ 2 T and gradually changes to $\sim$ B$_{a}^{1/3}$ below 1 T. At 30 K with B$_{m}$ = 0.2 T, resistivity $\sim$ B$_{a}^{2/3}$ for B$_{a} >$ 2 T and gradually changes to linear B$_{a}$ below 1 T. At 20 K with B$_{m}$ $\sim$ 2.4 T, similar linear variation of resistivity with B$_{a}$ was observed for B$_{a} >$ 3.5 T.

\begin{figure}
\includegraphics{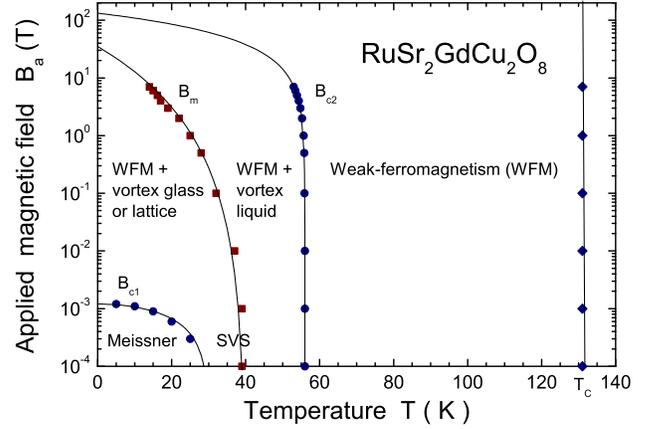}
\caption{\label{fig:epsart}  Full phase diagram B$_{a}$(T) of RuSr$_{2}$GdCu$_{2}$O$_{8}$.}
\end{figure}

\begin{figure}
\includegraphics{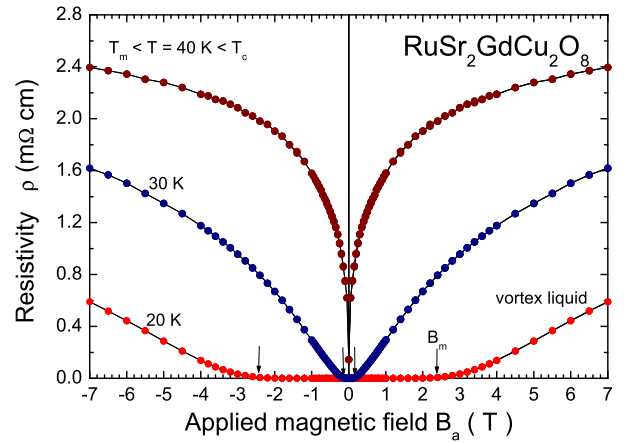}
\caption{\label{fig7}  Field dependence of magnetoresistivity $\rho$(B$_{a}$) for 
RuSr$_{2}$GdCu$_{2}$O$_{8}$ in the vortex state at 20 K, 30 K, and 40 K.}
\end{figure}

The last issue to be addressed is the depression of T$_{c}$ by small spontaneous Ru magnetic moments. The weak-ferromagnetic order is actually a canted antiferromagnetic order that can coexist with superconductivity. However, the observed T$_{c}$ of 56 K is too low as compared with 93 K for YBa$_{2}$Cu$_{3}$O$_{7}$ or 103 K for TlBa$_{2}$CaCu$_{2}$O$_{7}$. The depression of T$_{c}$ by small spontaneous magnetic moment can be partially recovered by substitution of nonmagnetic Cu ions at Ru site. For example, in the Ru$_{1-x}$Cu$_{x}$Sr$_{2}$GdCu$_{2}$O$_{8}$ system, T$_{c}$ onset up to 65 K for x = 0.1 and 72 K for x = 0.4 was reported.\cite{p26,p29}    

\section{conclusion}
The lower critical field with B$_{c1}$(0) = 12 G and T$_{0}$ = 30 K indicates the existence of a spontaneous vortex state (SVS) between 30 K and T$_{c}$ of 56 K. This SVS state is closely related with the weak-ferromagnetic order with a net spontaneous magnetic moment of $\sim$ 0.1 $\mu_{B}$ per Ru. The broad vortex liquid region observed above vortex melting line B$_{m}$(T) is also due to the coexistence and the interplay between superconductivity and weak-ferromagnetic order.        

\begin{acknowledgments}
This work was supported by the National Science Council of R.O.C. under contract No. 
NSC93-2112-M007-011. We thank Dr. B. N. Lin for helpful discussions.
\end{acknowledgments}

\newpage 

\end{document}